\documentstyle[12pt]{article}
\begin{document}
\centerline{\bf A Dynamical Approach to Fractional Brownian Motion}

\centerline{Riccardo Mannella}
\centerline{Dipartimento di Fisica dell'Universit\`{a} di Pisa}
\centerline{Piazza Torricelli 2, 56100 Pisa, Italy}

\centerline{Paolo Grigolini\footnote{Also at: Dipartimento di Fisica
dell'Universit\`{a} di Pisa, Piazza Torricelli 2, 56100 Pisa, Italy, and
Department of Physics of the University of North Texas,
P.O. Box 5368, Denton, Texas 76203}}
\centerline{Istituto di Biofisica del Consiglio Nazionale delle Ricerche}
\centerline{Via San Lorenzo 28, 56127 Pisa, Italy}

\centerline{Bruce J. West}
\centerline{Department of Physics, University of North Texas}
\centerline{P.O. Box 5368, Denton, Texas 76203}

\noindent
\section*{Abstract}
Herein we develop a dynamical foundation for fractional Brownian
Motion.
A clear relation is established between the asymptotic
behaviour of the correlation function and diffusion in a dynamical
system. Then, assuming that scaling is applicable, we establish a
connection between diffusion (either standard or anomalous) and the
dynamical indicator known as the Hurst coefficient. We argue on the
basis of
numerical simulations that
although we have been able to prove scaling only for "Gaussian"
processes, our conclusions may well apply to a wider class of systems.
On the other hand systems exist for which scaling might not
hold, so we speculate on the possible consequence on the various
relations
derived in the paper on such systems.

\newpage

\section{Introduction}

A stochastic process is in general characterized by two quantities, the
probability density describing the random nature of the fluctuations
and
the correlation function describing how a fluctuation at a given time
influences subsequent fluctuations. The statistics and the spectrum of
the
random variations are therefore independent quantities and both are
necessary to completely
describe a stochastic process. For example, a Gaussian distribution
may specify the statistics of a process, but the time dependence of the
variance, i.e. the width of the distribution, depends on the correlation
function or spectrum of the process.
Given that in many physical systems the Gaussian
distribution is a straightforward consequence of the Central Limit
Theorem,
investigators are often satisfied with measurements of the correlation
function to describe natural stochastic processes
particularly in the observation of large scale phenomena such as in
geophysics.
One such scientist who was most successful in this regard was Hurst,
who was particularly interested in problems of Hydrology and the Nile
river~\cite{it1,it2}.

Hurst~\cite{it1,it2,it3} developed a method called rescaled range
analysis, in which the span of a random process is devided by its
variance, resulting in a new variable that depends on the time length of
the data
record in a particularly interesting way. Let us define the time average
of the
random process $\xi(t)$ over the interval of time $\tau$: introducing
$t$, the discrete integer-valued time at which the observations are
recorded, and $\tau$, the total integer valued
time-interval considered, we have
\begin{equation}
\langle \xi \rangle_\tau = \frac{1}{\tau }\sum^\tau_{t =1} \xi (t)
\label{e1p1}
\end{equation}

Let us also define $X(t)$, the "accumulated departure" of $\xi(t)$ from
the mean
$\langle \xi \rangle_\tau$, i.e.
\begin{equation}
X (t,\tau ) \equiv \sum^t_{u = 1} \xi (u) - \langle \xi \rangle_\tau
\label{e1p2}
\end{equation}
so that the span of the process is defined by
\begin{equation}
S(\tau) = \max_{1 \le t \le \tau} X (t,\tau) - \min_{1 \le t \le \tau} X
(t,\tau)
\label{e1p3}
\end{equation}
Finally, let us also consider the
standard expression for the variance
\begin{equation}
V({\tau}) = \left[ \frac{1}{\tau}\sum^\tau_{t = 1} \left( \xi(t) -
\langle \xi \rangle_\tau \right)^2 \right]^\frac{1}{2}
\label{e1p4}
\end{equation}
The rescaled Hurst analysis consists in studying the properties of the
ratio
\begin{equation}
R (\tau) = S (\tau) / V (\tau)
\label{e1p5}
\end{equation}
He found that for the time records of over 850 phenomena $R$
is well described by the following empirical relation
\begin{equation}
R (\tau )= \left( \tau / 2 \right)^{H_H}
\label{e1p6}
\end{equation}
Here we refer to $H_H$ as the Hurst exponent (Hurst used the
symbol $K$ for this exponent).

Mandelbrot and co-workers~\cite{it4,it5,it6} modeled this heuristic
result
using the theory of fractional Brownian motion.
They showed that fractional Brownian motion would
provide an explicit statistical realization of (~\ref{e1p6}),
and that the theory would imply a reasonable
interpretation of the parameter $H_H$.
We emphasize that this important result supports the
interpretation of
natural phenomena in terms of fractal functions. However,
this
interpretation does not take into account the fact that fractal functions,
as
important as they may be, are still idealizations of natural phenomena.
These idealizations are not applicable to all time scales. In this
context we
mention the recent efforts to establish the physical foundation
of classical statistical mechanics using the concept of
chaos~\cite{it7,it8,it9}. These efforts rely
on there being a wide time scale separation between the microscopic
and the
macroscopic dynamical regimes.
The microscopic quantities of motion are valid on
a very short time scale, where the
conventional statistical arguments are inapplicable. Then, upon
increasing
of the time
scale considered, as a result of the action of chaos, the system of
interest
exhibits those statistical properties which are associated with
the
conventional ideas of a canonical distribution and Gaussian statistics.

The main purpose of the present paper is to provide a dynamical basis
for the
Hurst
rescaled range analysis. We show that the theory of Mandelbrot and
co-workers~\cite{it4,it5,it6} focuses on the asymptotic limit of the
dynamical processes considered herein.

There are a number of benefits arising from this change in
perspective. First of all, adding a "dynamical dimension" to the
Hurst
analysis, can be profitably used to quantitatively illustrate the
evolution
of a deterministic system away from a regular toward a totally
chaotic process, the
latter being, for many purposes, virtually indistinguishable from a
stochastic one.
Thus, the Hurst exponent may provide an independent means of
distinguishing
stochastic from chaotic motion~\cite{it10}.
Following this idea, we are tempted
to
speculate that the distinction between these two kinds of physical
processes
may be
merely a question of time scale.

Secondly we shall see that such a dynamical analysis, i.e., putting the
fractal
geometry of Mandelbrot into a dynamical setting, has the beneficial
effect
of
rigorously establishing a connection between the Hurst coefficient and
the behaviour of the autocorrelation functions at long times. We
link the coefficient to the existence of positive or negative tails
for the autocorrelation function of the random variable $\xi$,
\begin{equation}
C(t) = \langle \xi (0) \xi (t) \rangle
\label{e1p7}
\end{equation}
In molecular dynamics, dynamical processes characterized by a
long-time
regime with an inverse power-law correlation function
\begin{equation}
C(t) = \pm k/t^\alpha \; \; \; \; \; \; \mbox{for } \;\; t \rightarrow \infty
\label{e1p8}
\end{equation}
are denoted as {\em slow-decay} processes since the pioneering work
of Adler
and Wainwright~\cite{it11} where (~\ref{e1p8}) was first identified,
correlation functions have been the subject of an intense
debate~\cite{it12,it13}. From (~\ref{e1p8}) we note that the power
law
decay can have either a positive or a negative tail. We shall see that
the processes that Mandelbrot denotes as {\em persistent}, $H_H > 1/2$
in (~\ref{e1p6}), are connected by our theoretical analysis to a
positive tail, whereas those termed by him to be {\em antipersistent},
$H_H <
1/2$
in (~\ref{e1p6}), are connected by our theoretical analysis to
the existence of negative tails.

We support the analytic arguments presented in Section 2 and 3
with computer calculations
done for a substantial number of different dynamical systems.
The numerical results presented in Section 4
support the suggested dynamical approach to
the "geometrical" theory of Mandelbrot, but they also suggest that the
asymptotic time regime itself might be explored with different
mathematical
arguments, valid also for non-Gaussian statistics.

\section{Dynamical theory}
Let us now focus our attention  on the following equation of motion
\begin{equation}
\dot{x} = \xi
\label{e2p1}
\end{equation}
The Hurst coefficient (~\ref{e1p6}) was defined in terms of a
discrete time process so any dynamical representation such as
(~\ref{e2p1})
ought to be discrete as well. However, to connect the process with the
field of molecular dynamics, we adopt a continuous time
representation. The formal time integration of (~\ref{e2p1}) yields
\begin{equation}
x (t) = \int_0^t \xi  (t') d t' + x(0)
\label{e2p2}
\end{equation}

Thus the variable $x(t)$ undergoes a kind of motion with erratic
fluctuations induced by $\xi(t)$. Later on we shall define more precisely
the nature of such
"disordered" motion. For the time being we limit ourselves to the
conventional language of statistical mechanics. Thus, rather than
focusing on single
trajectories we shall study the mean values of quantities like $x^n (t)$.
We
make
the simplifying assumption that the erratic variable $\xi(t)$ fluctuates
around
zero. Thus $\langle x(t) \rangle = \langle x(0) \rangle$ where the
brackets denotes an average over an ensemble of realisations of
$\xi(t)$, as well as the distribution of initial state of $x(t)$.
We are therefore
obliged to study the mean value of $x^2 (t)$. By averaging $x^2 (t)$
over the fluctuations and initial conditions, we obtain
\begin{equation}
\langle x^2 (t) \rangle =  \int_0^t dt' \int_0^t dt'' \langle \xi (t') \xi
(t'')
\rangle
+ 2 \int_0^t dt' \langle \xi (t') x(0) \rangle +
\langle x^2 (0) \rangle
\label{e2p3}
\end{equation}
We make the assumption that the second term on the r.h.s. of
(~\ref{e2p3})
vanishes, hypothesizing no correlation between the
initial
value of $x$ and $\xi$. This assumption certainly holds true when
there exists a large time
scale separation between the dynamics of the {\em fast variable}  $\xi $
and that of the
{\em slow variable} $x$. However, this hypothesis must be used
with
caution in applications, which often refer to situations of slow
decay (~\ref{e1p8}), implying an exceptionally extended memory. We
then adjust the limits of the time integrals and obtain
\begin{equation}
\langle x ^2 (t) \rangle =
2\int_0^t dt' \int_0^{t'} dt'' \langle \xi (t') \xi (t'') \rangle +
\langle x^2 (0) \rangle
\label{e2p4}
\end{equation}

Under the assumption that the process $\xi (t)$ is stationary, i.e., its
moments
are independent of the time origin, so that
\begin{equation}
\langle \xi (t') \xi (t'') \rangle =\langle \xi (t'-t'') \xi (0) \rangle
\label{e2p5}
\end{equation}
we derive from (~\ref{e2p4}) the following integro-differential
equation of motion
\begin{equation}
\frac{d}{dt}\langle x^2 (t) \rangle = 2 \int_0^t \langle \xi (\tau) \xi (0)
\rangle d \tau
\label{e2p6}
\end{equation}
Clearly, the $\langle x^2 (t) \rangle$ appearing in the l.h.s.
of (~\ref{e2p6}) must be connected to the long-time diffusional regime
described by
\begin{equation}
\langle x^2 (t) \rangle = K t^{ 2 H_D}.
\label{e2p7}
\end{equation}
It is evident that the physical bounds on the possible values of $H_D$
are given by
\begin{equation}
0 < H_D < 1;
\label{e2p8}
\end{equation}
$H_D = 0$ defines the case of localization, which is the
lower limit of any diffusion process, and $H_D = 1$
obviously refers to the case of many uncorrelated deterministic
trajectories, with $x(t) - x(0)$ linearly proportional to time for each
of them. The bound $H_D < 1$ is a consequence of the fact that a
diffusional process cannot spread faster than a
collection of deterministic trajectories! Finally, the condition $H_D =
1/2$
is obtained for simple Brownian motion, where the variance increases
linearly with time.

On the other hand, using the definition of the correlation function
given in (~\ref{e1p7}), (~\ref{e2p6}) becomes
\begin{equation}
2 D(t) \equiv \frac{d}{dt}\langle x^2 (t) \rangle = 2 \int_0^\infty C(t')
dt'.
\label{e2p9}
\end{equation}
We can now show that, using (~\ref{e2p9}), the deviation of $H_D$
from the conventional diffusion prediction $H_D = 0.5$ can be explained
if
the correlation function $C(t)$ exhibits a slow decay. The joint use
of (~\ref{e1p8}), (~\ref{e2p7}) and (~\ref{e2p9}) leads to the following
long-time prediction
\begin{equation}
\frac{d^2}{dt^2}\langle x^2 (t) \rangle = 2 H_D ( 2 H_D -1 ) K t^{2
H_D-2} \sim 2 C(t) = \pm \frac{2k}{t^\alpha}
\label{e2p10}
\end{equation}
having assumed that the long-time limit of the correlation function
$C(t)$ is
dominated by the inverse power law of (~\ref{e1p8}). The positive
(negative) sign
refers to the case of the solid (dashed) line in Fig. 1.

{}From (~\ref{e2p10}) we determine that $H_D$ and $\alpha$
satisfy the relation
\begin{equation}
H_D = 1 - \alpha/2
\label{e2p11}
\end{equation}
obtained by matching the time dependences. It is also clear
from the coefficients in (~\ref{e2p10}) that
$H_D > 1/2$ implies a
positive long-time correlation, whereas $H_D < 1/2$ implies a negative
long-time correlation.
Let us now summarize the result of this simple theoretical analysis,
with
an eye to Fig. 1:
\begin{eqnarray*}
\mbox{Case exemplified by the solid line:} \; & 1 > H_D > 1/2; \;\;\;\; 1
> \alpha > 0 \\
\mbox{Case exemplified by the dashed line:} \; & 0 < H_D < 1/2; \;\;\;\; 2
> \alpha > 1 \\
\end{eqnarray*}
Thus, we see that the solid line correlation function of Fig. 1 leads
to a superdiffusive
behavior ranging from the standard diffusion $(H_D = 1/2)$ to the
ballistic
behavior $(H_D = 1)$. The dashed line correlation function of the same
figure leads to a
subdiffusive behavior ranging from the standard diffusion to no motion
at all.

It must be stressed that for superdiffusive correlation functions
(~\ref{e2p9}),
$D(\infty)  = \infty$, whereas in the
case of subdiffusive correlation functions, $D(\infty)$ is finite or even
vanishing.
In this latter case we obtain what we define as {\em classical Anderson
localization}. At early times in the diffusional process the mean
square value of $x(t)$ increases. Then when the negative part of the
correlation function $C(t)$ becomes important, the rate of diffusion
decreases. When the negative tail completely compensates for the
positive
part of the relaxation process, the rate of diffusion virtually vanishes.
At this late time stage further diffusion is rigorously prevented and the
diffusing
particle becomes localized. Processes of this kind have recently been
discovered~\cite{it14} and the theory presented here affords a
remarkably
straightforward explanation of them. It is interesting that such
processes
should admit such a simple interpretation.

We point out that in the simple theory presented in this section, the
only significant assumptions made are the stationary property
of (~\ref{e2p5})
and the absence of correlation between  $x(0)$ and $\xi$. No
assumption
was made
on the nature of the statistics of the stochastic process $x$
except that it has a finite correlation function. In the next
section we shall rederive (~\ref{e2p6}) under the more
restrictive assumption that the process $\xi$ and therefore $x$
is Gaussian. We also note that
the standard case $H_D = 0.5$ is compatible with almost any kind of
relaxation
process. The only condition to fulfill is that the correlation function
$C(t)$ is square integrable over the time interval . Thus, if we exclude
the case $H_D = 0.5$, we must invoke
an inverse power law decay to explain the behavior
given by (~\ref{e2p7}). This is so because (~\ref{e2p7})
implies the existence
of "stationary"
behavior and which, in turn, implies the existence of
an inverse power-law decay. A
power-law decay is the only way of "killing" the possibility itself of
defining a time scale, and this, in turn, is a condition essential to
explain the "stationary" nature of the diffusion regime
of (~\ref{e1p7}) with $H_D \ne 0.5$.
What do we mean by "stationary"? Upon increase of the time scale
considered,
the diffusion process is increasingly dependent on the tail of the
correlation function, until it becomes totally dominated by the
inverse power-law
tail. In this regime, since the power-law decay of
the correlation function implies no time scale
is dominant, the diffusion process becomes
stationary. The concept of a stationary diffusion process can be easily
expressed by referring to the Hurst coefficient rather than to
$H_D$. We shall see that $H_H$ is time dependent, and that it usually
reaches a
stationary value for $t\to \infty$. This definition of "stationary"
diffusion behavior implies that the diffusion coefficient $H_D$ can be
identified with the asymptotic value of the Hurst coefficient $H_H$. We
shall see that this is frequently the case even if we shall only be able
to {\em rigorously} prove it for Gaussian statistics.

\section{A Fokker-Planck treatment}

The next step is to determine the connection of the Hurst coefficient
$H_H$
with the diffusion
coefficient $H_D$. To do this we need to derive suitable expressions for
the quantities appearing in (~\ref{e1p5}), for any given dynamical
system. In part we follow the strategy of Mandelbrot:
the idea is that for a white Gaussian
process it is simple to carry out the theoretical analysis, hence it is
only necessary to find, for a given dynamical system, the corresponding
"Gaussian" approximation in the appropriate "reduced" time scale.

The details of the approach are given in reference~\cite{it15}.
For the given dynamical system we want
to study, we replace the dynamical equations with the
following equation of motion for the probability density
\begin{equation}
\frac{\partial}{\partial t}\rho (x,\xi ,\Gamma ;t) = L \rho (x,\xi
,\Gamma ;t).
\label{e3p1}
\end{equation}
Here $\Gamma$ stands for the entire set of variables necessary to
describe
the time evolution of $\xi $. The "Liouville-like" operator $L$ is divided
into two parts as follows,
\begin{equation}
L = L_I + L_B
\label{e3p2}
\end{equation}
where the "interaction" part determined by (~\ref{e2p1}) yields and the
phase space operator
\begin{equation}
L_I = - \xi \frac{\partial}{\partial x},
\label{e3p3}
\end{equation}
and $L_B$ defines the time evolution of the distribution of the variables
$\xi $ and $\Gamma$,
determined by the dynamics of the corresponding set of variables. It is
not necessary to define the explicit form
of the latter operator since it depends in detail on the specific
problem studied.

We now use the Zwanzig projection approach~\cite{it16}, which consists
in
integrating the total distribution $\rho$ over the degrees of freedom
that
are not of interest to us
\begin{equation}
\sigma (x,t) = \int d\xi d\Gamma \rho (x, \xi , \Gamma )
\label{e3p4}
\end{equation}
We apply this projection approach following the perturbation
prescriptions of~\cite{it14}, assuming that $L_I$ in (~\ref{e3p3})
is a "weak" perturbation. Using this basic
assumption, after some algebra described in detail in~\cite{it15},
we arrive at the result
\begin{equation}
\frac{\partial}{\partial t}\sigma (x,t) = \Xi (t)
\frac{\partial^2}{\partial x^2}\sigma (x,t),
\label{e3p5}
\end{equation}
where
\begin{equation}
\Xi (t) = \int_0^t C(\tau ) d \tau
\label{e3p6}
\end{equation}

If (~\ref{e3p5}) was used to determine the time evolution of
$\langle x^2 (t) \rangle$ i.e., multiply (~\ref{e3p5}) by $x^2$
and integrate over $x$, it would reduce to (~\ref{e2p9}).
In this sense (~\ref{e2p9})
and (~\ref{e3p5}) are equivalent. Unfortunately, the two equations are
not truely equivalent
because (~\ref{e2p9}) is obtained without making any
assumption regarding the statistics of $x$,
whereas (~\ref{e3p5}) is really the result of a
second-order perturbation treatment, equivalent to assuming that the
statistical process $x$ is Gaussian.

If now we rescale the time as follows
\begin{equation}
t^{*} = t^{2 H_D},
\label{e3p7}
\end{equation}
(~\ref{e3p5}) can be written as
\begin{equation}
\frac{\partial }{\partial t^{*}}\sigma (x; t^{*}) = \Phi \left( t^{*}
\right)\frac{\partial ^2}{\partial x^2}\sigma (x; t^{*})
\label{e3p8}
\end{equation}
where, of course,
%
%
\begin{equation}
\label{e3p9}
\Phi \left( t^{*} \right)=\frac{1}{2 H_D}{t^{*}}^{\frac{1}{2H_D}-1}
\int_0^{t^{* \frac{1}{2H_D}-1}} \frac{dt'^{*}}{2 H_D}
{t'^{*\frac{1}{2H_D}-1}} C \left( t'^{*} \right)
\end{equation}
After some algebra it can be shown that
%
%
\begin{equation}
\label{e3p10}
\lim_{t^{*} \rightarrow \infty} \Phi \left( t^{*} \right) = K
\end{equation}
where $K$ is a finite constant. It is evident that in the
asymptotic time limit (~\ref{e3p8}) becomes a standard (time
independent) Fokker-Planck equation, and the statistical process
defined in terms of the scaled variable
%
%
\begin{equation}
\label{e3p11}
y \equiv \frac{x}{\sqrt{K t^{*}}},
\end{equation}
becomes a Gaussian process with the distribution
%
%
\begin{equation}
\label{e3p12}
p (y) = \frac{1}{\sqrt{2 \pi }}\exp \left( - \frac{y^2}{2} \right).
\end{equation}
This result means that the random function $x(t)$ is proportional to
${t^{*}}^{1/2}$ or,
in the original time scale, to $t^{H_D}$. Following
Mandelbrot~\cite{it4}, we are then
led to identify $H_D$ with $H_H$.

This is the central theoretical result of this paper. In the
asymptotic limit the dynamical process described by (~\ref{e3p1})
becomes
the fractional Brownian motion of Mandelbrot. However, in the short
time
regime the process can be substantially different from such a stochastic
process. This is clearly illustrated by the numerical results of the
following section.

We note that $H_D$ is the coefficient appearing in
(~\ref{e2p7}).
However, to be able to identify this coefficient with $H_H$ it is
necessary
that we go through the Fokker-Planck equation of (~\ref{e3p5}), and
this implies making a second-order approximation, absent in the
derivation of (~\ref{e2p7}).
In the next section we show, using numerical methods that $H_H$
is very close in value to $H_D$ even in cases when (~\ref{e3p5})
should not be valid. This suggests the possibility of establishing the
property $H_H = H_D$ without using the second-order approximation
necessary for the derivation of (~\ref{e3p5}).

\section{Numerical simulations}

	We now present numerical simulations of a number of different
dynamical systems, to which the Hurst analysis is then applied.
Comments on the
correspondence of $H_D$ and $H_H$ are discussed alongside the
numerical
data. For clarity, the different models are presented within separate
subsections.

\subsection{Microscopic dynamics described by the Langevin equation}

The easiest way to create a stochastic process $\xi$ with an exponential
correlation function is through the Langevin equation
%
%
\begin{equation}
\label{e4p1}
\dot{\xi } = - \gamma \xi (t) + f(t)
\end{equation}
where $f(t)$ is Gaussian white noise. Clearly, the relevant time scale
here is given by $1/\gamma$, and we expect that for times larger than
$1/\gamma$, we should find $H_H=0.5$. This is confirmed (see Fig. 2)
by the
numerical simulations: the expected asymptotic regime is reached at
shorter times for the case of larger $\gamma $.

A similar model is given by
the multidimensional Langevin equation
%
%
\begin{eqnarray}
\label{e4p2}
\dot {\xi } &=& w (t) \nonumber \\
\dot {w}  &=& -\Gamma w (t) - \Omega^2 \xi (t) + F(t)
\end{eqnarray}
In the case $\Gamma \gg \Omega$, (~\ref{e4p2}) becomes
indistinguishable from (~\ref{e4p1})
with $\gamma = \Omega^2/\Gamma$. Again, as soon as the time scale
considered is larger than the typical time scale of the system, we should
have $H_H=0.5$. However, here a problem arises. It is possible
to consider either the time under the correlation function
or the time over which the "quasi energy" in (~\ref{e4p2}) looses
correlation:
these two times are in general very different. By inspection of Fig.
3, it is clear that numerical simulations done for different $\Omega^2$
and the
same $\gamma $ show a similar asymptotic behavior more or less in
the same time
region (compare the solid line and the dotdashed line in Fig. 3):
whereas, when a different $\gamma $ value is considered (compare the
solid and
dashed lines), the asymptotic regime sets in at different times.
This seems to imply that the relevant time scale is the time over which
the energy looses correlation.

Due to the linearity of (~\ref{e4p2}), the Gaussian statistics of the
stochastic force is transmitted to $\xi $ and hence to $x$. Thus, the
Central
Limit Theorem is fulfilled and the asymptotic behavior must be
characterized by $H_H = H_D = 0.5$.  The parallelism between $R$ and
$\langle x^2 (t) \rangle^{1/2}$ of
Fig. 4 shows that the transient behavior corresponds exactly to the time
it takes for the second moment to reach the stationary condition,
corresponding to standard diffusion. In the case of (~\ref{e4p2}) this
transient time is a complex function of the parameters $\Gamma $ and
$\Omega $, which we
do not discuss here. But of course standard diffusion sets
in over time scales which are connected to the time taken by the
quasi-energy to decorrelate, hence justifying the behaviour
shown in Fig. 3.

\subsection{Bistable stochastic motion}

The next system we consider involves
the motion of a stochastic particle in
a bistable potential. The dynamics can be cast in the dimensionless form
%
%
\begin{equation}
\label{e4p3}
\dot {\xi } = \xi - \xi^3 + f(t)
\end{equation}
where $f(t)$ is a stochastic Gaussian white force [see (~\ref{e4p1})] of
intensity $D$.
The Hurst rescaled range
analysis is applied to the variable $\xi$. In the limit of small
noise intensities $D$ the system (~\ref{e4p3}) is characterized by two
different time scales: one is the time scale of the inter-well relaxation
time
($T_R$); the other one is the time scale of the intra-well dynamics
($T_K$,
related to the Kramers rate for the system). For $\tau \gg T_K$
the system should
behave as a dichotomous random process, hence $H_H=0.5$. Also, if
applicable, in the range $T_R \ll \tau \ll T_K$ we expect $H_H=0.5$: this
is
because before
the Hurst analysis is able to single out the dichotomous random process
(the behaviour for $\tau \gg T_K$ ), the condition $T_R \ll \tau$ implies
that, for the
relevant $\tau$ values, the dynamics is similar to that of
(~\ref{e4p1}). This
is confirmed by Fig. 5, where we show the result for $R (\tau)$ obtained
by digitally simulating (~\ref{e4p3}). For the simulations we use
$D=0.1$, which
yields a $T_K$ of approximately 30, and a $T_R$, determined by
the coefficients appearing
in the force in (~\ref{e4p3}), of order 1. It is possible to note that for
larger $\tau$ values $H_H$ approaches the correct value $0.5$, and that
for
$T_R < \tau < T_K$, $H_H$ is
smaller than it is, say, in the region $\tau \sim T_K$. Unfortunately,
the condition $ T_R \ll \tau \ll T_K$ is
only weakly satisfied by our choice of the parameter $D$ and it is not
possible
to observe exactly $H_H=0.5$ in the range of intermediate $\tau$.

\subsection{Noisy Lorenz model}

Another model to which we have applied the Hurst
rescaled range analysis is the Lorentz
model perturbed by Gaussian white noise~\cite{it17}. The
perturbed Lorenz model is
described by the stochastic differential equations
%
%
\begin{eqnarray}
\label{e4p4}
\dot\xi & = & \sigma \left(\xi - y\right) + f (t) \nonumber \\
\dot{y} & = & r \xi  - y - \xi z \\
\dot{z} & = & -  z + \xi y \nonumber
\end{eqnarray}
where $f(t)$ is a Gaussian white process with intensity $D$. We set
$\sigma = 8/3$, $b=10$ and $r=126.5$, a set of parameters for which
the Lorenz model is known to be periodic, as long as $D=0$. The chosen
value of the control parameter $r=126.5$ is roughly in the middle of a
small
periodic island within an $r$ region for which the Lorenz model
displays
chaos. The addition of a small stochastic force (i.e., $D \ne 0$) "kicks" the
system out of the periodic region, and leads to the observation of
chaos~\cite{it17}.
The transition periodic/chaotic motion as function of the noise
intensity $D$ is very smooth, and chaos, defined by a
positive Lyapunov exponent,
is observed for $D > 10^{-5}-10^{-4}$.

Applying the Hurst rescaled range
analysis to the variable $\xi $ of the Lorenz system, we
expect that $H_H$ will vary from around zero (the value for periodic
motion) to
0.5 (diffusive dynamics) for increasing noise. Also, the $\tau$ values for
which the asymptotic behaviour is observed should become smaller as
$D$ is
increased. The result of numerical simulations of (~\ref{e4p4}) is shown
in
Fig. 6. Note that in Fig. 6 we have plotted $R(\tau)/\tau^{1/2}$
versus $\tau$, to more clearly show
the asymptotic behaviour. For all curves, $R(\tau)$ increases for
small $\tau$, up to $\tau \sim 1$, which is the period of the noiseless
Lorenz
system
for the chosen parameters. In the case of very small noise intensities
(full and
dotted line), we have that for larger $\tau$, $R(\tau)$ goes like
$\tau^0$;
behaviour
typical of periodic motion. When we increase the noise intensity (small
dashes line), after an initial decrease, $R(\tau)$ increases: it
eventually
leads to $H_H=0.$5, for still larger $\tau$ values, but on the $\tau$
range shown the
asymptotic behaviour is not yet established. For this $D$ value,
we remark, that the Lorenz model is only very weakly chaotic. Finally,
for even larger noise intensities (large dashes line) the departure of
$R(\tau)$ from the noiseless curve takes place at yet smaller $\tau$
values, and
the
expected asymptotic behaviour ($H_H=0.5$, on the figure, represented
by an
horizontal line) for $R(\tau)$ is clearly identified.

\subsection{A two-dimensional potential}
We next study the deterministic motion of a particle in the
two-dimensional potential
%
%
\begin{equation}
\label{e4p5}
V(\xi ,y) = \cos \left(\xi +y \sqrt{3} \right) + \cos \left(\xi -y
\sqrt{3} \right) + \cos \left( 2 y / \sqrt{3} \right)
\end{equation}
which defines an infinite lattice of triangular symmetry (see~\cite{it18}
for more
details). The equations of motion we have integrated have the form
%
%
\begin{eqnarray}
\label{e4p6}
\dot\xi & = & v \nonumber \\
\dot{v} & = & -\frac{\partial V (\xi ,y)}{\partial \xi } \nonumber \\
\dot{y} & = & w \\
\dot{w} & = & -\frac{\partial V (\xi ,y)}{\partial y } \nonumber
\end{eqnarray}

It is known that when the energy of the system is
efficiently small the
particle moves from the bottom of each triangular
cell, only occasionally wandering from cell to cell. As a consequence, the
self-diffusion coefficient shows a peculiar behaviour as the energy is
changed.

It is clear that in principle this motion could lead to anomalous
diffusion: in particular, as the energy is decreased the "periodic" motion
within each cell should become more and more dominant in the
dynamics of
the particle. Obviously, the "random" diffusion from cell to cell is still
in place, hence we only expect some weak departure from $H_H=0.5$ as
the
mechanical energy is decreased. By inspection (see Fig. 2 in~\cite{it18})
it is
clear that the autocorrelation function of the velocity $v$ has a visible
negative tail (which is not at all surprising, remembering the
importance
of the periodic motion within each cell). We then expect that as the
energy is decreased $H_H$ should take on values smaller than 0.5. This
is
confirmed by the Hurst rescaled range
analysis applied to the variable $ \xi $, and shown in Fig. 7.

As clear from the discussion in the previous section, the fundamental
question is whether the coefficient $H_H$ should be
related to $H_D$ in situations of anomalous diffusion. We compare the
quantity $R(\tau)$ (Hurst analysis) and
$\langle v^2(\tau) \rangle^{1/2}$ (see Section 2) in Fig. 8: we used an
energy value equal to -0.90,
for which the dynamics is supposedly anomalous ($H_H = 0.43$). The
clear
parallelism between the two curves at large times establishes that the
diffusion is indeed anomalous as suggested by the Hurst analysis.

\subsection{Standard map}

We now present the results obtained in a discrete model, i.e. for the
standard map
%
%
\begin{eqnarray}
\label{e4p7}
x_{t+1} &=& x_t + \frac{K}{2 \pi} \sin \theta_t \nonumber \\
\theta_{t+1} &=& \theta_t + x_{t+1} \;\;\;\; (\mbox{mod} \;\;\; 2 \pi )
\end{eqnarray}
The standard map is very convenient to test our interpretation of the
Hurst rescaled analysis
analysis: it has been recently shown~\cite{it19} that for
appropriate
parameter values within the chaotic regime anomalous diffusion should
arise.

According to~\cite{it19}, the anomalous diffusion is caused by chaotic
orbits
sticking to critical tori encircling accelerator mode islands.
For this reason the correlation function $C(t)$ should have the power
law dependency of (~\ref{e2p10}). We studied the map for the same
values
considered in~\cite{it19}, i.e., for $K=3.86$, 6.4717, 6.9115 and 10.053:
the
dynamical variable we considered for the analysis is the quantity
$\xi_t=x_{t+1}-x_t$. A standard diffusion behaviour is expected for
$K=3.86$ and
10.053, and an anomalous diffusion behaviour for the other two $K$
values.
That this is qualitatively the case is clearly shown in Fig. 9, where we
have plotted $R(\tau)$ normalized to $\tau^{1/2}$ versus
$\tau$ for the different $K$'s: as
expected, the curve is an horizontal line (standard diffusion) for
$K=3.86$
and 10.053. We would now like to understand whether for the
anomalous
diffusion case we have some correspondence between the Hurst
rescaled range analysis and
the numerical work of~\cite{it19}.  In the case $K = 6.9115$,
in~\cite{it19}
it is reported that theory and numerical simulations lead to
%
%
\begin{equation}
\label{e4p8}
\alpha \sim \frac{2}{3}
\end{equation}
Let us insert (~\ref{e4p8}) into (~\ref{e2p11}). Adopting the
notation of~\cite{it19}, i.e., $\zeta \equiv 2 H_D$, we obtain
%
%
\begin{equation}
\label{e4p9}
\zeta \sim \frac{4}{3}=1.3333333...
\end{equation}
According to~\cite{it19} this prediction fits very well the result of the
numerical calculation on diffusion.

We remark once more that for $K = 3.86$ and 10.053, values for which
the authors of~\cite{it19}
observe a standard diffusion we obtain, with a very high
degree of accuracy $H_H=0.5$;
and where ($K=6.9115$ and 6.4717) anomalous diffusion is predicted
we obtain a value of $H_H$ significantly different from
$H_H=0.5$. Moreover, for $K = 6.9115$ we obtain $\zeta  = 1.2330$,
to be compared to $\zeta  = 1.3333$ from (~\ref{e4p9}).

Table 1 summarizes the situation. Note that the  fourth column,
$\zeta_{sim}$
denotes the result of the numerical simulation~\cite{it5} (the only value
there reported corresponds to $K = 6.9115$) and that the last column,
$\zeta(H_H)$
reports the values of $\zeta$ corresponding to the value of $H_H$
evaluated
numerically (third column).

\section{Conclusions}

It must be pointed out that the Mandelbrot analysis, leading to $H_H =
H_D$,
is essentially based on the Central Limit Theorem, assumed
to be valid even when anomalous diffusion occurs. The adoption of the
Fokker-Planck treatment of Section 3 leads to the following time
dependence of the $x$-distribution:
%
%
\begin{equation}
\label{e5p1}
\sigma (x;t) = \frac{1}{\left( 2 \pi K t^{2H} \right)^{1/2}}\exp \left( -
\frac{x^2}{4 K t^{2H}} \right)
\end{equation}
It is then evident that the moment $\langle x^n(t) \rangle$ rescales in
time as
$t^{2H}$. Since
the Hurst rescaled range
analysis refers to a quantity with the same dimension of $x$ it is
evident that it leads to
%
%
\begin{equation}
\label{e5p2}
H_H = H = H_D
\end{equation}

The functional form (~\ref{e5p1}) suggests that in general,
after an initial transient, the probability distribution
$\sigma (x;t)$ should perhaps be described by the equation
%
%
\begin{equation}
\label{e5p3}
\sigma (x;t) = \frac{1}{t^\beta} F \left( \frac{x}{t^\beta} \right)
\end{equation}
If the rescaling of (~\ref{e5p3}) applied, we would have
that indeed in general $H_H=H_D$.

Let us now briefly discuss some possible forms of $F$.
There are three possible conditions:
\begin{description}
\item[(i)] $\beta=1/2$, $F$ is a Gaussian function of its argument. This
is
the standard diffusion process.
\item[(ii)] $\beta \ne 1/2$, $F$ is a Gaussian function of its argument.
This is the
fractional Brownian motion process.
\item[(iii)] $\beta \ne 1/2$, $F$ is not a Gaussian function of its
argument.
Note that this occurs for a L\'evy stable process~\cite{it20}.
\end{description}

It must be pointed out, however, that from a physical point of view it is
hard to imagine a diffusion process with a deterministic origin agreeing
with (~\ref{e5p1}), and thus falling under case (ii), even in the case
$H > 0.5$. The reason is that
as established by the theoretical analysis of Section 2, this
anomalous behavior comes from an anomalously slow correlation
function,
namely the correlation  function of (~\ref{e1p8}) with $\alpha < 1$.
In this physical situation, there is no hope to realize (~\ref{e5p1})
as an effect of the Central
Limit Theorem: the original process must be already Gaussian! In other
words, if there existed Gaussian statistical processes leading to the slow
decay of $C(t)$, then the anomalous diffusion would be compatible with
the time
rescaling of (~\ref{e5p1}). In our opinion, this is the physical nature of
the
fractional Brownian motion of Mandelbrot. It is the long-time
asymptotic
limit of a Gaussian process with an anomalously slow correlation
function.

We think that this situation might occur in statistical mechanics when
the source of the Brownian motion, the statistical process $x$,
refers to a physical condition characterized by a large number of
degrees
of freedom. However, in the last few years, there have been attempts to
build statistical mechanics on chaos, without  the
joint action of a very large number of degrees of
freedom~\cite{it7,it8,it9}.
In this
physical situation $x$ is a non-Gaussian statistical process and the
Gaussian nature of diffusion stems from the action of the Central Limit
Theorem. If the process is not Gaussian, but it is fully chaotic, then the
correlation function is exponential or, more generally, characterized by
a well-defined time scale.

In the special case when chaos and ordered motion coexist, however,
the
dynamical behavior of the system becomes extremely more complex,
and a
correlation function with an inverse
power law might occur. This implies the
breakdown of the time scale separation between diffusion and
microscopic
dynamics, and the consequent breakdown of the Central Limit Theorem
itself. In this physical condition (~\ref{e5p1}) cannot apply.

Are there processes rescaling according to (iii), without implying the
Gaussian assumption? We think that if an anomalous diffusion exists,
then
it is quite probable that it belongs to the class (iii). We are convinced
that some of the processes examined in Section 4 belong to class (iii). If
the rescaling in (~\ref{e5p3}) with $\beta \ne 1/2$ holds true, then
we conclude
immediately that $H_H = \beta$. However, this special condition raises
the
intriguing question of whether or not $H_H = H_D$, in this case. The
dynamical
realization of the diffusion process is expressed by (~\ref{e2p1}).
Let us assume the $X$
is the maximum possible value of $\xi$. It is then evident that at the
time $N$
the $\xi$ distribution must be contained between  $x_m$ and  $- x_m$,
with $x_m = NX$.
Now, let us imagine that there are theoretical reasons to expect that the
$x$ distribution is characterized by long tails with an
inverse power law $1/x^m$. It is
then evident that the rescaling of (~\ref{e5p3}) cannot apply to the
whole space.
This might generate a discrepancy between $H_H$ and $H_D$. Let us
assume, for
simplicity, that $x_m = A t$. In such a case we get a rescaling of the
same
kind as (~\ref{e5p3}) only for $|x_m/t| < A$. Thus the moments of the
distribution
rescale with a power law different from that leading to the time
rescaling
of (~\ref{e5p3}).

We wonder if a possible discrepancy between the two coefficients might
be
derived using the data already available for the standard
map~\cite{it19}. According
to~\cite{it19}, the distribution rescales as in (~\ref{e5p3}), with
$\beta = 3/5$ for
$K = 6.9115$. However, this distribution is truncated at the value $|y |=
|x/t^\beta|= 1$. If we assume that $H_H$ is determined by the rescaling
of (~\ref{e5p3}),  we obtain
%
%
\begin{equation}
\label{e5p4}
2 H_H = 2 \beta = 6/5 = 1.2
\end{equation}
thereby suggesting that the discrepancy between the numerical value
$2 H_H = 1.23330$, obtained in this paper, and the numerical value $2
H_D
= 1.3333333$,
determined numerically by the authors of~\cite{it19}, might be due not
to the
inaccuracy of the direct calculation of $H_D$ in~\cite{it19} (notice that
the
calculation of $H_H$ is expected to more accurate than that of $H_D$),
but it
might rather depend on the breakdown of the condition $H_H = H_D$,
due
to the non-Gaussian character of the distribution $F$ of (~\ref{e5p3}).

We shall address these questions in further investigations. For the time
being
we must limit ourselves to saying that the Hurst rescaled range
analysis seems to be an
efficient numerical technique to explore how a dynamical system
approaches
its long-time asymptotic limit, or, equivalently, which is the short time
dynamics of that asymptotic idealization referred to by
Mandelbrot~\cite{it4,it5,it6}
as fractional Brownian motion.

\section*{Acknowledgments}
We warmly thank professor Jaczek Kowalski on many illuminating
discussions
on the possible difference between $H_H$ and $H_D$. This work was
partially supported by the EC under contract n. SC1-CT91-0697 (TSTS).
One of us (B.J.W.) also thanks the Naval Air Warfare Center for partial
support of this research [contract No. N62269-92-C-0548] and the
Office of Naval Research [contract No. 73191].

\newpage
\section*{Table caption}

Table summarising the comparison between $H_H$ from numerical
simulations (third column) as function of the parameter $K$ (first
column). The power of the correlation function
tail, from the simulations, is shown under the heading $\alpha$.
For the definition of $\zeta$ see text. Note that $\zeta_{sim}$ is the
value computed in~\cite{it18}, and that $\zeta (H_H)$ is computed from
our $H_H$ values.

\newpage
\section*{Figure captions}

\begin{itemize}
\item Figure 1: Typical slow decaying correlation functions with a
positive tail (solid line) and a negative tail (dashed line).

\item Figure 2: the Hurst rescaled range
analysis applied to the Brownian motion of~\ref{e4p1}.
The labels indicate the resulting Hurst function $R(\tau)$, for different
$\gamma$. The
straight lines are best fits to the appropriate power law.

\item Figure 3: The Hurst rescaled range
analysis is applied to the Brownian motion of~\ref{e4p2}.
The parameters values are: full line, $\Omega^2=10$, $\gamma=1$;
dashed line
$\Omega^2=1$, $\gamma=10$; and
dot-dashed line $\Omega^2=1$, $\gamma=1$.

\item Figure 4: Comparison between the Hurst rescaled
analysis (dashed line) and diffusion of the variable $x$
($=\langle x^2(\tau)\rangle^{1/2}$, see Section 2, solid line) in the
system described by~\ref{e5p2}. Parameter values chosen are
$\Omega^2=1$,
$\gamma=10$. Note the parallelism between the different curves at
large $\tau$ times.

\item Figure 5: The Hurst rescaled range
analysis applied to the system described by~\ref{e4p3}:
for $D\simeq 0.1$ chosen, $T_K\approx 30$, and $T_R \sim 1$ (see
text). The
boxed
numbers are the best fit $H_H$ values evaluated in the region around
the arrow
head.

\item Figure 6: The Hurst rescaled range
analysis is applied to the Lorenz model (~\ref{e4p4}). The
curves
are drawn for increasing noise intensities (solid line, $D=0$, dotted line,
barely visible in the bottom left corner,$D=10^{-6}$, small dashes line,
$D=10^{-5}$,
large dashes line, $D=10^{-4}$). The quantity plotted is $R(\tau)$
normalized
to $\tau^{1/2}$ versus
$\tau$, and in case of $H_H=0.5$ we should have an horizontal line.

\item Figure 7: The Hurst rescaled range
analysis is applied to the two-dimensional model
of~\ref{e4p5} and~\ref{e4p6}. From top to bottom the energy (in
round brackets $H_H$) has the
value $-0.60\ (0.49),\ -0.80\ (0.50),\ -0.85\ (0.49),\ -0.90\ (0.43)$
and -0.95 (0.39). We plotted the quantity $R(\tau)$ normalized
to $\tau^{1/2}$ versus
$\tau$, and in case of $H_H=0.5$ we should have an horizontal line.

\item Figure 8: Comparison between the Hurst rescaled range
analysis (solid line) and
diffusion ($=\langle v^2(\tau) \rangle^{1/2}$, see Section 2, dashed line)
for the model
of~\ref{e4p5} and~\ref{e4p6}, and for a mechanical energy equal
to -0.90 ($H_H=0.43$).
The parallelism between the curve proves the relevance of the Hurst
analysis in the calculation of the diffusion at large times.

\item Figure 9: The Hurst rescaled range is applied
analysis of the standard map,~\ref{e4p7}. The quantity $R(\tau)$
normalized to $\tau^{1/2}$ versus
$\tau$ is plotted, and in case of $H_H=0.5$ horizontal line results. The
different curves refer to different values of the parameter $K$ (see
text).
We have: solid line, $K=3.86$; small dashes line, $K=6.4714$; large
dashes
line, $K=6.9115$; and dot dashed line, $K=10.053$. Standard diffusion is
expected for $K=3.86$ and $K=10.053$.
\end{itemize}

\newpage
\centerline{\bf Table I}
\vspace{2.5truecm}
\centerline{
\begin{tabular}{|| l | l | l | l | l ||}  \hline\hline
$K$ & $\alpha$ & $H_H$ & $\zeta_{sim}$ & $ \zeta\left( H_H \right) $\\
\hline
$10.053$ &$ 0.9483 \pm 0.0058$ & $0.50317 \pm 0.0004$ & &
$1.0624$ \\
\hline
$6.9115$ & $0.80188 \pm 0.00888$ & $0.61665 \pm 0.00081$ &
$1.3333$ &
1.2333\\
\hline
$6.4717$ & $0.88778 \pm 0.00854$ & $0.57652 \pm 0.00066$ & &
$1.15304$\\
\hline
$3.86$ & $0.86957 \pm 0.00443$ & $0.50134 \pm 0.000307$ & &
$1.0268$ \\
\hline\hline
\end{tabular}
}


\begin{thebibliography}{18}
\bibitem{it12}
The literature on this topic is huge. For this reason, we limit ourselves
to
quoting only the recent theoretical paper of reference~\cite{it13}, which
provides the first microscopic derivation of an inverse
power law. The reader can find
there also an extended set of references on the subject of slow decay.
\end{thebibliography}
\end{document}